\documentclass[final]{eps}

\usepackage{graphicx}
\usepackage{times}
\usepackage{natbib}

\volumenumber{100}
\publishyear{2011}
\frompage{\pageref{firstpage}}
\topage{\pageref{finalpage}}

\def\Alfven{Alfv\'{e}n~}

\def\lesssim{\; \buildrel < \over \sim \;}

\shorttitle{Takeru K. Suzuki: Solar Wind Evolution.}

\title{Solar Wind and its Evolution }
\author{Takeru K. Suzuki$^1$}
\affiliation{Department of Physics, Nagoya University, 
$^1$Furo-cho Nagoya, 606-8602, Japan }
\received{xxxx xx, 2011}\revised{xxxx xx, 2011}\accepted{xxxx xx, 2011}
\abstract{
By using our previous results of magnetohydrodynamical simulations 
for the solar wind from open flux tubes, I discuss how the solar wind 
in the past is different from the current solar wind. 
The simulations are performed in fixed one-dimensional super-radially 
open magnetic flux tubes by inputing various types of fluctuations 
from the photosphere, which automatically determines solar wind properties 
in a forward manner.  
The three important parameters which determine physical properties of the solar 
wind are surface fluctuation, magnetic field strengths, and the configuration 
of magnetic flux tubes. 
Adjusting these parameters to the sun at earlier times in a qualitative sense, 
I infer that the quasi-steady-state component of the solar wind in the past 
was denser and slightly slower if the effect of the magneto-centrifugal force 
is not significant. I also discuss effects of magneto-centrifugal force and 
roles of coronal mass ejections. 
}

\keywords{atmosphere -- MHD --- planets ---solar corona --- solar wind 
--- waves}

\begin{document}
\label{firstpage}
\maketitle
\copyrighttext{}

\section{Introduction}

Young solar-type stars are generally very active: X-ray flux is up to
$\sim 1000$ times larger than %a middle-aged main sequence star, 
the present-day sun \citep{gud97,gud04}, and the X-ray temperature is 
also higher \citep{rib05,tel05}.  
%Furthermore, observations of classical T-Tauri stars show magnetic 
%field strength, $\gtrsim$ kG \citep{jk07}, 
Observations of young main sequence stars show very strong magnetic field 
strengths with an order of kG or even larger \citep[][see also Donati 
\& Landstreet 2009 for recent review]{dc97,sb99,saa01}, which are 
much stronger than the average strength of 1-10 G of the current sun.
Young stars are generally fast rotators and strong magnetic fields are 
generated by strong differential rotation through dynamo activities of 
magnetoconvection \citep[e.g.,][]{bru04}. As time goes on, a star loses 
its angular momentum through magnetic stellar wind \citep{wd67} and 
the magnetic activities become weak \citep[e.g.,][]{ayr97}. 
%Although the mechanism is not well understood, people think that stellar 
%rotation is a key because the stellar activities show nice correlation with 
%stellar rotation rate \citep{ayr97}, which is a decreasing function 
%on time as a result of the loss of angular momentum through magnetic stellar 
%wind \citep{wd67}. Differential rotation in the stellar 
%interior is inferred to play a role in this link between stellar 
%rotation and surface activity through dynamo activities of magnetoconvection 
%\citep[e.g.,][]{bru04}.

Solar wind is hot plasma emanating from the Sun, and the mass loss 
rate amounts to $\sim 10^{12}$g s$^{-1}$ 
($2\times 10^{-14} M_{\odot}$yr$^{-1}$) at present. 
As inferred from high activity of young solar-type stars, 
I expect that the solar wind was stronger at earlier times. 
\citet{wod02,wod05} observed asterospheres of low-mass stars by the 
Hubble space telescope and determined the mass loss rates.   
The estimated mass loss rates show a decreasing trend with time, 
%consistent with the trend of the stellar activities discussed above, 
except for very young stars, thought there is still a large scatter in the 
data, partly because the observed stars range from very low-mass to near 
solar-mass stars\footnote{Although the observed stars include not only G-type 
(solar-analogue) stars but also lower mass K- and M-type stars, it is 
worth discussing trends because the mechanism of mass loss of these stars 
are supposed to be similar; the stellar wind is 
accelerated by wave and turbulent motions associated with magnetic field 
driven by fluctuation in a surface convective zone.}.  
At the very early epoch, the mass loss rate seems to be saturated at 
$\sim$ 100 times of the present solar value, but some very young stars show 
lower mass loss rates than the present Sun. 

The solar wind roughly consists of two components. 
The first component is called fast solar wind with %, of which the speed is 
speed of 700 -- 800 km s$^{-1}$ at the earth orbit. 
Fast solar wind is more steady and streams out from coronal 
holes which generally corresponds to open magnetic flux tube regions. 
The other component is slow solar wind with speed $\lesssim 400$ km s$^{-1}$.  
%which show more complex characters. 
Slow wind is more complex and transient, and in most cases comes from lower 
latitude regions. Recent HINODE observations show that some portions of the 
slow wind appear to be originating from open flux tubes near active regions 
\citep{sak07,ima07,hrr08}, while there are still debates of other sources 
for slow wind.  

%Although it is not categorized in the solar wind in a narrow sense, 
As other types of magnetized plasma from the sun, coronal mass ejections 
(CMEs) also are supposed to affect atmospheres of planets especially at 
early times. 
%In addition to the solar wind from open coronal holes, coronal 
%mass ejection (CME) is also magnetized plasma ejected from the sun. 
%In terms of evolution of atmospheres of planets at early epochs, CMEs, in 
%addition to solar wind from open flux tubes, are supposed to play important 
%roles. 
In this paper, however, I mainly focus on solar wind from open magnetic 
flux tubes, and briefly mention effects of CMEs afterward. 
Our group has carried out forward numerical modeling of the solar wind by using 
direct magnetohydrodynamical (MHD) simulations, in which we can directly test 
the response of the solar atmosphere to the surface fluctuations and properties 
of magnetic fields. By using the results of the simulations 
with parameters suitable for the younger sun, I discuss the solar wind 
evolution.

\section{Simulation Model}
In this section, I briefly describe the simulations of the solar wind. 
For detail, please refer to \citet[][SI06, hereafter]{si06}. 
In order to cover the region with huge density contrast from 
$\rho = 10^{-7}$g cm$^{-3}$ at the photosphere to 
$\rho \sim 10^{-21}$g cm$^{-3}$ 
at the outer simulation boundary located at $\approx 0.1$ astronomical 
unit (AU), simple one dimensional (1-D) open flux tubes are adopted. 
The effects of super-radial expansion of flux tubes are incorporated by 
taking into account an expansion factor in the conservation of 
magnetic flux of radial ($r$) component, $B_r$:
%The simulation regions are from the photosphere ($r=1R_{\odot}$) with  
%density, $\rho = 10^{-7}$g cm$^{-3}$, to $65R_{\odot}$ (0.3AU) or $\simeq 
%20 R_{\odot}$ ($\simeq$0.1AU), where $R_{\odot}$ is the solar radius.  
\begin{equation}
B_r r^2 f(r) = {\rm const.} ,
\end{equation}
where $f(r)$ is a super-radial expansion factor. 
%This super-radial expansion is a result that 
Most of the solar surface is covered by closed magnetic loop structure. 
Then, open flux tubes rapidly open above these loops; $f$ is introduced 
to consider this effect. 
The same function as in \citep{kh76} is adopted for $f(r)$ (see SI06 
for detail).

In SI 06 we input the transverse fluctuations of the field line by the 
granulations at the photosphere. 
Amplitude, $\langle dv_{\perp} \rangle$, at the 
photosphere is chosen to be compatible with  
the observed photospheric velocity amplitude $\sim 1$km s$^{-1}$ 
\citep{hgr78}. 
SI06 tested various types of spectra; in this paper I discuss 
the results of the power spectra in proportion to $1/\nu$, where 
$\nu$ is frequency.
%, which excite \Alfven waves. 
%In this paper we only show results of linearly polarized perturbations 
%with power spectrum proportional to $1/\nu$, where $\nu$ is frequency (see
%SI06 for circularly polarized fluctuations with different spectra).
The surface fluctuations generate upgoing \Alfven waves which contribute 
to the acceleration of the solar wind in upper regions. 
At the outer boundaries, a non-reflecting condition is imposed for all the MHD 
waves, which enables us to carry out 
simulations for a long time until quasi-steady state solutions are obtained  
without unphysical wave reflection. 

SI06 dynamically treat the propagation and dissipation of the waves and the 
heating and acceleration of the plasma by solving 
ideal MHD equations with the relevant physical processes, including the sun's
gravity, radiative cooling, and thermal conduction \citep[][SI06]{szi05}.  
I do not take into account stellar rotation, which may be important 
for very young stars. 
For the initial condition, we assume a static atmosphere with temperature 
$T=10^4$K in order to see whether the atmosphere is heated up to coronal 
temperature and accelerated to accomplish the transonic flow. 
From $t=0$, the transverse fluctuations are injected from the 
photosphere and continue the simulations until the quasi-steady states 
are achieved.

\section{Solar Wind Response to Surface Condition}
By injecting fluctuations from the photospheric surface, the initially 
cool and static atmosphere is effectively heated and accelerated by the 
dissipation of the generated upgoing \Alfven waves. 
The sharp transition region which divides the cool chromosphere with 
$T\sim 10^4$K and the hot corona with $T\sim 10^6$K is formed owing to a 
thermally unstable region around $T\sim 10^5$K in the radiative cooling 
function \citep{LM90}. 
The hot corona streams out as the transonic solar wind 
\citep[see][\& SI06 for detail]{szi05}. 

The heating and acceleration of the solar wind plasma in inner heliosphere 
results from the dissipation of \Alfven waves.
In the simple 1-D treatment, \Alfven waves mainly dissipate via nonlinear 
mode conversion; slow MHD waves are nonlinearly generated from outgoing 
\Alfven waves and the slow MHD waves are damped by shock formation 
as a result of steepening of wave shape. The shocks also heat up surrounding 
plasma, which play a central role in the heating of the solar wind plasma. 
Magnetic pressure associated with the \Alfven waves decreases with height as 
a result of the successive dissipation process, which directly pushes the 
plasma outward (momentum input) in addition to gas pressure. 

The young sun is more active than the present-day sun, and therefore, I expect 
that surface fluctuations are stronger. The properties of the magnetic fields  
are also supposed to be very different. 
I discuss how properties of the solar wind are affected by fluctuation 
amplitudes and magnetic fields at the solar surface. 

\subsection{Dependence on Surface Fluctuation Amplitude}
\begin{figure}[t]
  \vspace*{2mm}
  \begin{center}
    \includegraphics[width=0.8\linewidth]{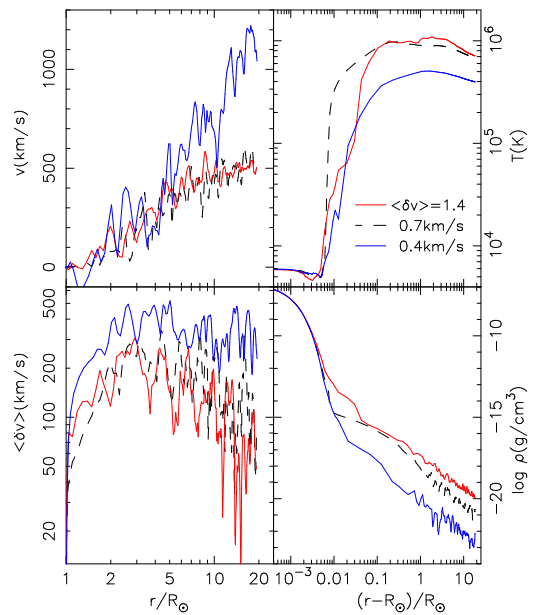}
  \end{center}
  \caption{Structures of corona and solar wind for different 
$\langle dv_{\perp}\rangle=$0.4 (blue), 0.7 (black dashed), 
and 1.4 km s$^{-1}$ (red). I plot solar wind speed, 
$v_r$(km s$^{-1}$) (top left), temperature, $T$(K) (top right), 
density in logarithmic scale, $\log (\rho({\rm g\; cm^{-3}}))$ 
(bottom right), and rms transverse velocity, 
$\langle dv_{\perp}\rangle$(km s$^{-1}$) (bottom left). 
Each variable is averaged with respect to time during 28min (the longest 
wave period considered).}
  \label{fig:dspsw}
\end{figure}

\begin{table*}[t]
\begin{tabular}{|c|c|c|c|c|}
\hline
$\langle dv_{\perp}\rangle$ & $T_{\rm max}$ & $n_{\rm p,1AU}$ 
& $v_{\rm 0.1AU}$ & $(n_{\rm p}v)_{\rm 1AU}$\\ \hline
1.4(km/s) & $1.2\times 10^6$(K) & 50(cm$^{-3}$) & 500(km/s) 
& $2.5\times 10^9$(cm$^{-2}$s$^{-1}$) \\ \hline
0.7(km/s) & $1.0\times 10^6$(K) & 5(cm$^{-3}$) & 550(km/s) 
& $2.8\times 10^8$(cm$^{-2}$s$^{-1}$)\\ \hline
0.4(km/s) & $0.5\times 10^6$(K) & 0.05(cm$^{-3}$) & 1200(km/s) 
& $6\times 10^6$(cm$^{-2}$s$^{-1}$) \\ \hline
\hline
\end{tabular}
\caption{Summary of model parameters and results. The first column 
shows the input amplitudes of the surface fluctuation. The second 
column presents the maximum temperatures in the simulation regions.
The third column shows proton number density at 1 AU, which is calculated 
from $n_p$ at the outer boundary of the simulation by assuming the conservation 
of the mass flux ($n_p v r^2 f=$const.) with constant speed from 0.1 AU 
to 1 AU. 
The fourth column shows the solar wind speeds at 0.1 AU ($\simeq$ the outer 
boundary of the simulation region). 
The fifth column shows mass flux of the solar wind at 1 AU, 
which is the product of the third and fourth columns. }
\label{tab:depdv}
\end{table*}

Figure \ref{fig:dspsw} shows the response of the solar wind plasma to 
the surface fluctuations with different amplitudes, $\langle dv_{\perp} 
\rangle = 1.4$km s$^{-1}$, 0.7km s$^{-1}$ and 0.4km s$^{-1}$. % cases. 
The intermediate case ($\langle dv_{\perp} \rangle = 0.7$km s$^{-1}$) is 
a reference case which explains the present-day solar wind.

The maximum temperature of $\langle dv_{\perp} \rangle = 0.4$km s$^{-1}$ 
case is $\simeq 5\times 10^5$K (Table \ref{tab:depdv}), which is cooler 
than the usual corona. The density is lower than the fiducial
case by 1-2 orders of magnitude because the sufficient mass cannot be supplied 
to the corona by the chromospheric evaporation owing to the low 
temperature; the evaporation is drastically suppressed as $T$ decreases, 
since the conductive flux ($F_c \propto T^{5/2}\frac{dT}{dr}$) sensitively 
depends on $T$. 
As a result, the mass flux ($\rho v_r$) becomes much lower, by a factor of 
$\sim$50, than that of the present solar wind. 

On the other hand, the larger $\langle dv_{\perp} \rangle=1.4$km s$^{-1}$ case 
gives larger coronal density, although the coronal temperature is only slightly
higher than the temperature of the reference case. As a result of 
the dense corona, the mass flux of the solar wind is $\sim$10 times larger 
(Table \ref{tab:depdv}).
This case shows complicated structure of temperature and density near the 
solar surface.
The temperature starts to increase from a deeper location around 
$r\simeq 0.005R_{\odot}$ than the other cases. Thanks to this,  
the decrease of the density is slower (larger pressure scale height) so that 
the density around $r=1.01R_{\odot}$ is two orders of magnitude larger than 
that of the reference case with $\langle dv_{\perp} \rangle = 0.7$km s$^{-1}$. 
However, the temperature decreases slightly instead of showing a 
monotonical increase; it cannot go over the peak of the radiative cooling 
function at $T\simeq 10^5$K \citep{LM90} because the radiative loss is 
efficient, owing to the large density. 
The temperature again increases from $r\simeq 1.03R_{\odot}$ and above 
there the corona forms. 
The coronal density is larger than the that of the reference case as explained 
above. 

A striking feature is that the mass flux of solar winds sensitively 
depends on the input wave amplitude at the solar photosphere. 
Since the injected energy is proportional to the square of $dv_{\perp}$, 
the difference of the input energy between the small amplitude case 
($\langle dv_{\perp} \rangle = 0.4$km s$^{-1}$) and the large amplitude 
case ($\langle dv_{\perp} \rangle = 1.4$km s$^{-1}$) is $\simeq$ 12. 
However, the output mass flux of the large amplitude case is $\sim$ 400 
times larger than the mass flux of the small amplitude case.

The sensitive behavior on the surface amplitude can be understood 
by reflection and nonlinear dissipation of \Alfven waves (SI06). 
%(Figure \ref{fig:dpwam2}). 
Because the heating is smaller, the temperature is lower in the smaller 
$\langle dv_{\perp} \rangle$ cases. Then, the scale height becomes smaller 
and the density decreases rapidly. The \Alfven speed, $v_{\rm A}(=B_r
/\sqrt{4\pi\rho})$, changes more rapidly due to the rapid decrease of density 
and the wave shape is largely deformed, which enhances 
the reflection. When the input wave energy decreases, a positive feedback 
operates; a smaller fraction of the energy can reach the coronal region. 
As a result, the density and mass flux of the solar wind becomes much 
smaller than the decreasing factor of the input energy. 

In addition to the effect of the wave reflection, the nonlinear dissipation 
of the \Alfven waves also plays a role in the sensitive dependence on 
the input wave energy (SI06). When the input wave energy 
becomes smaller, the density becomes smaller as explained above. 
Then, the nonlinearity of \Alfven wave, $d v_{\perp}/v_{\rm A}$, 
decreases, not only because the amplitude, $d v_{\perp}$, is smaller but 
also because the \Alfven speed, $v_{\rm A}(\propto 1/\sqrt{\rho})$, is larger. 
Therefore, the \Alfven waves do not dissipate and the heating is 
reduced, which further decreases the density; this is another type 
of positive feedback, which also results in the sensitive dependence 
of the density on the input wave energy.

\subsection{Magnetic Fields}
At earlier epochs the sun is expected to have possessed stronger magnetic 
field than today, as inferred from observations of young stars 
\citep{dc97,sb99}
%\citep{jk07}. 
%In terms of driving solar wind, 
The configuration of magnetic fields as well as field strength plays an 
essential role in determining the physical solar wind conditions. 
%If one models open magnetic flux tubes by super-radially open 1-D tubes as 
%in this paper, 
In the 1-D treatment as adopted in this paper, a super-radial expansion 
factor, $f$, determines the geometry of flux tubes. 
\citet{ws90,ws91} showed that the solar wind speed at 
$\sim$ 1AU is anti-correlated with $f$ from their long-term 
observations. This can be naively %na\"{i}vely 
understood by energetics
consideration; since $f$ directly determines an adiabatic loss of plasma in 
flux tubes, solar wind plasma in largely expanding (larger $f$) flux tubes 
undergoes larger energy loss. 

By the comparison of the outflow speed obtained by interplanetary 
scintillation measurements with observed photospheric field strength 
\citet{kfh05} %and \citet{hak05} 
have found 
that the solar wind velocity is better correlated with surface magnetic 
field strength, 
divided by the expansion factor, $B_{r,0}/f$: 
They claim that $B_{r,0}/f$ is a better index for the speed 
of solar wind than individual $1/f$ or $B_{r,0}$.
The correlation between $B_{r,0}/f$ and solar wind speed is well explained 
by \Alfven waves in expanding flux tubes \citep{suz06}.  Because the energy 
flux of input \Alfven waves is proportional to $B_{r,0}$, 
the positive correlation of solar wind speed with $B_{r,0}$ is quite natural. 
The negative correlation with $f$ reflects the adiabatic loss 
of \Alfven waves in expanding tubes as explained above.  
%In other words, the correlation with $B_{r,0}/f$ reflects \Alfven waves 
%in expanding flux tubes. 

One can understood the correlation more specifically from nonlinear dissipation 
of \Alfven waves as well.
The nonlinearity of the \Alfven waves, $\langle d
v_{\perp}\rangle /v_{\rm A}$
is controlled by $v_{\rm A} \propto B_r \propto B_{r,0}/f$ in
the outer region where the flux tube is already super-radially open.  
Wave energy does not effectively dissipate in the larger $B_{r,0}/f$ 
case in the subsonic region because of relatively small nonlinearity, and 
more energy remains in the supersonic region. 
In general, energy and momentum inputs in the supersonic region result in 
higher wind speed, while those in the subsonic region raise the mass flux 
($\rho v_r$) of the wind by increasing density \citep{lc99}.  
This indicates that the solar wind speed is positively correlated with 
$B_{r,0}/f$.

\section{Evolution of Solar Wind}
I have described based on our numerical simulations that the solar wind from 
open magnetic flux tubes is mainly controlled by the three parameters: 
surface fluctuation amplitude, $\langle dv_\perp \rangle$, radial magnetic 
field strength at the photosphere, $B_{r,0}$, and super-radial expansion 
factor, $f$, of flux tubes. Thus, if I can determine these parameters for 
the young sun, I can estimate the properties of solar wind at that time. 

\subsection{Surface Fluctuation Amplitude}
I can speculate that $\langle dv_\perp \rangle$ was larger at earlier times 
from circumstantial evidences.  
First, the rotation of the sun should be faster at earlier times because 
of loss of angular momentum with time. Hence, in addition to surface 
convection, interior differential rotation will lead to stronger surface 
fluctuations. From observational facts, younger stars shows higher X-ray 
activities as discussed in the introduction section, which also anticipates 
larger $\langle dv_\perp \rangle$.  
An important issue is that a small increase of $\langle dv_\perp \rangle$ leads 
to a large increase of mass flux of solar wind. 

\subsection{Magnetic Field --$B_{r,0} \& f$--}
$B_{r,0}$ is also supposed to be larger than today as implied from    
observations of young stars \citep{dc97,sb99}. 
Rossby number, $R_0$, which is defined as the ratio of stellar rotation period 
to convective turn-over time, is an important index which describes  
stellar magnetic activities \citep{noy84}. 
Surface magnetic field strength, $B_{\rm ph}$, 
multiplied by a magnetic areal filling factor, $\cal{F}$, is well-correlated 
with $R_0^{-1}$. Stars with small $R_0 < 0.1$, which correspond 
to fast rotating young stars, have $B_{\rm ph} \cal{F} =$ 1-10 kG 
\citep{saa01}. In some stars $\cal{F}$'s are obtained and the values are 
50-70\%, which shows that $B_{\rm ph}$ is an order of 10 kG in these stars.

If all the magnetic flux is open to the interplanetary space, I can set 
$f=1/\cal{F}$; in reality, however, closed loop structure is more dominant, 
and then, it is expected that $f \gg 1/\cal{F}$.
I speculate that super-radial expansion, $f$, of the young sun was 
larger as discussed from now, although it is not simple to pin down 
the specific value.   
Recently, the configuration of magnetic field has been observed 
in a number of low-mass stars \citep[e.g.][]{don08}. 
Fast rotating solar mass stars possess non-axisymmetric poloidal components
with substantial toroidal fields \citep[see Fig.3 of ][]{dl09}. 
This implies that the field configuration is quite complicated, and is very 
different from ordered dipole structure. 
In comparison with 11-year periodicities of present-day solar activity 
\footnote{I would like to note that the 11-year duration is modulated with 
time depending activity even in last several hundred years \citep{miy08}}, 
the properties of magnetic field of the earlier sun resemble the condition of  
the present-day solar maximum rather than the solar minimum. 
% which is observed during the solar minimum. 

%observations of T-Tauri stars 
%%show magnetic field strength of $\gtrsim$ kG 
%\citep{jk07}. 
%%, which is $\sim 100$ times larger than the present Sun in terms of 
%%the average field strength.
%Super-radial expansion, $f$, or more generally configuration of 
%magnetic fields, of the young Sun is quite uncertain compared to the 
%other two parameters. We here try to get some informations from the present 
%Sun. 

%periodic activities of the current Sun gives us some hints. 
%As is well knows, the Sun exhibits 11-year periodicities of its activities, 
%whereas the period seems slightly change with time \citep{miy08}. 
During the solar minimum periods, the solar wind consists of fast wind which 
is from mid- to high-latitude regions, and slow wind from lower-latitude 
regions. On the other hand, during the solar maximum periods, most of 
the regions are occupied by transient slow wind \citep{mcc08}. 
At the solar maximum magnetic fields are dominated by a dipole component, 
while at the solar minimum higher multi-pole moments become more important 
showing more complex field configuration \citep{hak05}. 
Because the surface is mostly covered with closed magnetic fields during the 
solar maximum, the field strength in the outer regions is not as strong as 
that at the solar minimum although the field strength at the photosphere 
is stronger. 
This fact shows that $f$ is larger during the solar maximum. 
The speed of the solar wind during the solar maximum is slower, 
which is consistent with the observational trend \citep{kfh05},  
because of larger adiabatic loss in open magnetic flux tubes.  

%Turning back to the younger sun, the condition of the magnetic fields is 
%supposed to be more similar to the maximum phase of the present-day Sun. 
%If this speculation is correct, 
Based on these considerations, namely (i) the magnetic field structure of 
the young sun more resembles the current solar maximum than the solar 
minimum, and (ii) open flux tubes show larger expansion during the solar 
maximum because larger surface area is occupied by closed structure, I infer 
that open magnetic flux tubes of the young sun 
have much larger $f$. I speculate that, even though the magnetic field 
strength at the footpoints is stronger, $B_{r,0}/f$, is smaller at earlier 
time. 

\subsection{Solar Wind in the Past}
I would like to discuss how the properties of solar wind change in the 
past based on the simulation results. 
%within the framework of our simulations here. 
I have stated that $\langle dv_\perp \rangle$ was larger and $B_{r,0}/f$ 
was smaller at earlier times. %If they are correct, 
These conditions imply considerably 
denser but slightly slower solar wind in the past, although it is quite 
difficult to quantitatively determine the physical condition. 
%Then, the speed of solar wind from young Sun is slower than today. 
%While it is difficult to discuss speeds of solar wind at earlier 
%times in a quantitative manner, the speed is roughly at least about 
%the level of the present slow wind ($\sim 300 - 400$ km s$^{-1}$). 
%This is because the wind speed should be an order of the escape velocity 
%($\sim 600$ km s$^{-1}$) from the Sun, and slower wind cannot flow out 
%because of gravity. 

%Summing up these informations, we speculate that solar wind at earlier 
%times would be denser but slightly slower than today, 

\subsection{Limitations}
In this paper I neglect the effect of magneto-centrifugal force in driving 
winds \citep{wd67,mes68}. Under the typical solar condition, 
if the rotation period is 4 days 
(6-7 times faster than the present sun) or less, magneto-centrifugal force 
plays an important role and the wind speed becomes significantly higher 
\citep{new80}. 

Recently, \citet{hj07} calculated the evolution of solar wind by considering 
megneto-centrifugal force. They concluded that the solar wind in the past 
was faster but not so much denser than today, which are different from my 
conclusion.   
\citet{hj07} adopted the density at a 'reference point', 1.1 
times of the stellar radius, in a parameterized way; they assume power-law 
dependence on stellar rotation rate.
Since the basal density strongly constrains the mass loss rate 
\citep[see, e.g. section 3 of][]{lc99} and the adopted dependence in their 
standard case is 
rather weak, the derived mass loss rate shows weak dependence with time.   
In a strict sense, the density at the coronal base, accordingly 
the density at the reference level, is determined by the energy balance 
among heating, thermal conduction, and radiative cooling (SI06; see 
also Suzuki 2007 for different types of stars). 
The base densities of young active stars are larger than those of quiescent 
stars, and the mass loss rates are supposed to be significantly larger. 
In this case the wind speed becomes slower because more mass needs to 
be pushed away. The same tendency on the coronal base density 
is also reported by \citet{ste11} who performed 3D MHD simulations for 
a young sun.  

%Frankly speaking, 
It is not still possible to discuss the wind speed 
in a quantitative sense at the moment. My conclusion of the slightly slower 
wind in the past might be corrected because magneto-centrifugal 
force could be significant. On the other hand, I suppose that the wind 
speed is not so high as estimated by \citet{hj07} bacause the coronal base 
density should larger at that time than their assumption. 
Accordingly, the density of the solar wind could be larger 
than those calculated in \citet{hj07}.

In this paper I focus on quasi-steady-state solar wind from open magnetic 
flux regions,  
%, which is rather quasi-steady-state component. 
and do not consider CMEs. 
% which are dynamical events probably triggered by 
%magnetic reconnections, also contributes to the plasma supply to the
%interplanetary regions. 
While the quasi-steady-state component dominantly 
contributes to the total mass loss from the present-day sun than CMEs, at early 
times CMEs might play a more important role because of stronger magnetic 
activities. 
%We discussed that the speeds of the solar wind from open flux tubes is 
%supposed to be slower than today. 
From the fossil record on the lunar surface, it is inferred that the speed of 
solar wind was faster in the past \citep{rh80,new80}. I infer that 
the fossil record comprises effects of both quasi-steady state component 
and CMEs. 
If strong CMEs dominantly contribute to the mass loss, the bulk speed of 
plasma flow from the sun would be recognized as higher. 
I speculate that the fossil inprint on the moon is connected with high 
CME activities at early epochs. 
%
%There are still a lot of open questions remained.   

\section{Summary}
Based on our previous results of MHD simulations, I discuss the evolution 
of the quasi-steady-state component of solar wind from open flux tubes. 
The properties of the solar wind are determined by the three parameters, 
surface fluctuation amplitude, magnetic field strength, and expansion of 
open flux tube.  
Referring to observational data of young low-mass stars, I suppose that 
the surface amplitude was larger and the ratio of field strength to 
flux tube expansion was smaller.  
Following this speculation, the solar wind in the past was dense and 
slightly slower than today, while if the effect of magneto-centrifugal force 
is taken into accout, the speed of the solar wind might be higher 
at very early epoch than my conclusion. 
CMEs are also supposed play a more important role in the past than today 
because of higher magnetic activities. 
In order to understand the effects of solar 
magnetized plasma on the atmospheres of planets, we need to understand the 
evolution of CMEs as well as the evoltuion of quasi-steady-state solar 
wind.

%\subsubsection{Subsubsection}\strut

\acknowledgments{acknowledgment}
This work was supported in part by Grants-in-Aid for 
 Scientific Research from the MEXT of Japan, 19015004, 20740100, 
 22864006, and Inamori Foundation.
The author thanks two anonymous reviewers for many constructive 
suggestions to improve the paper.  
Guest editor M. Yamauchi thanks two anonymous reviewers in
evaluating this paper. (Received February 3, 2011; Revised April 6,
2011; Accepted April 11, 2011).
%%\lastpagecontrol{20cm}

%\begin{references}\frenchspacing

\email{T.K.Suzuki (e-mail: stakeru@nagoya-u.jp)}
\label{finalpage}
\lastpagesettings

\begin{thebibliography}{99}

\bibitem[Ayres (1997)]{ayr97}
Ayres, T. R., Evolution of the solar ionizing flux, 
\textit{J. Geophys. Res.}, \textbf{102}, 1641 - 1652, 1997

\bibitem[Brun et al.(2004)]{bru04}
Brun, A., S., Miesch, M. S., Toomre, J., 
Global-Scale Turbulent Convection and Magnetic Dynamo Action in the 
Solar Envelope, \textit{Astrophys. J.}, \textbf{614}, 1073 - 1098, 2004

\bibitem[Donati \& Collier Cameron (1997)]{dc97}
Donati, J.-F. \& Collier Cameron, A., Differential rotation and magnetic 
polarity patterns on AB Doradus, \textit{Mon. Not. Roy. Astron. Soc.}, 
\textbf{291}, 1 - 19, 1997

\bibitem[Donati et al.(2008)]{don08}
Donati, J.-F. et al., Large-scale magnetic topologies of early M dwarfs, 
\textit{Mon. Not. Roy. Astron. Soc.}, \textbf{390}, 545 - 560

\bibitem[Donati \& Landstreet (2009)]{dl09}
Donati, J.-F. \& Landstreet, J. D., Magnetic Field of Nondegenerate Stars, 
\textit{Ann. Rev. Astron. Astrophys.}, \textbf{47}, 333 - 370

\bibitem[G\"udel (2004)]{gud04}
G\"udel, M. X-ray astronomy of stellar coronae, \textit{Astron. 
Astrophys. Rev.}, \textbf{12}, 71 - 237, 2004

\bibitem[G\"udel et al.(1997)]{gud97}
G\"udel, M., Guinan, E. F., Skinner, S. L., 
The X-Ray Sun in Time: A Study of the Long-Term Evolution of Coronae 
of Solar-Type Stars, \textit{Astrophys. J.}, \textbf{483}, 947 - 960, 1997

\bibitem[Hakamada et al.(2005)]{hak05}
Hakamada, K., Kojima, M., Ohmi, T., Tokumaru, M, Fujiki, K., 
Correlation between Expansion Rate of the Coronal Magnetic Field 
and Solar Wind Speed in a Solar Activity Cycle
\textit{Sol. Phys.}, \textbf{227}, 387 - 399, 2005 

\bibitem[Harra et al.(2008)]{hrr08}
Harra, L. K. et al. Outflows at the Edges of Active Regions: Contribution 
to Solar Wind Formation?, \textit{Astrophys. J. Lett.}, \textbf{676}, 
L147 - L150, 2008

\bibitem[Holweger et al.(1978)]{hgr78}
Holweger, H., Gehlsen, M., \& Ruland, F.:
Spatially-averaged properties of the photospheric velocity field, 
\textit{Astron. Astrophys.}, \textbf{70}, 537 - 542, 1978 

\bibitem[Holzwarth \& Jardine (2007)]{hj07}
Holzwarth, V. \& Jardine, M., Theoretical Mass Loss Rates of Cool 
Main-sequence Stars, \textit{Astron. Astrophys.}, \textbf{463}, 11 - 21, 
2007

\bibitem[Imada et al.(2007)]{ima07}
Imada, S. et al., Discovery of a Temperature-Dependent Upflow in the 
Plage Region During a Gradual Phase of the X-Class Flare., 
\textit{Pub. Astron. Soc. Japan}, \textbf{59}, S793 - 799, 2007

\bibitem[Johns-Krull (2007)]{jk07}
Johns-Jrull, C. M., The Magnetic Fields of Classical T Tauri Stars, 
\textit{Astrphys. J.}, \textbf{664}, 975 - 985, 2007

\bibitem[Kojima et al.(2005)]{kfh05}
Kojima, M., K. Fujiki, M. Hirano, M. Tokumaru, T. Ohmi, and K. Hakamada, 
Solar Wind properties from IPS observations, 
\textit{The Sun and the heliosphere as an Integrated System},
 Giannina Poletto and Steven T. Suess, Eds. 
Kluwer Academic Publishers, 147 - 181, 2005

\bibitem[Kopp \& Holzer(1976)]{kh76}
Kopp, R. A. \& Holzer, T. E., Dynamics of coronal hole regions. I 
-- Steady polytropic flows with multiple critical points, 
\textit{Sol. Phys.}, \textbf{49}, 43 - 56, 1976 

\bibitem[Lamers \& Cassinelli(1999)]{lc99}
Lamers, H. J. G. L. M. \& Cassinelli, J. P., \textit{Introduction to 
Stellar Wind}, Cambridge, 1999

\bibitem[Landini \& Monsignori-Fossi(1990)]{LM90}
Landini, M. \& Monsignori-Fossi, B. C.:
The X-UV spectrum of thin plasmas, \textit{Astron. Astrophys. Supp.}, 
\textbf{82}, 229 - 260, 1990 

\bibitem[McComas et al.(2008)]{mcc08}
McComas, Ebert, R. W., Elliot, H. A., Golestein, B. E., Cosling, J. T., 
Schwadron, N. A., Skoug, R. M., 
Weaker solar wind from the polar coronal holes and the whole Sun., 
\textit{Geophys. Res. Lett.}, \textbf{35}, L18103

\bibitem[Mestel (1968)]{mes68}
Mestel, L., Magnetic Braking by a Stellar Wind - I., \textit{Mon. Not. Roy. 
Astron. Soc.}, \textbf{138}, 359 - 391

\bibitem[Miyahara et al.(2008)]{miy08}
Miyahara, H., Yokoyama, Y., Masuda, K., 
Possible link between multi-decadal climate cycles and periodic reversals 
of solar magnetic field polarity, \textit{Earth. Planet. Sci. Lett.}, 
\textbf{272}, 290 - 295

\bibitem[Newkirk (1980)]{new80}
Newkirk, Gordon Jr., Solar Variability on Time Scale of 10$^5$ years to 
10$^{9.6}$ years, \textit{Proc. Conf. Ancient Sun}, 293 - 320, 1980 

\bibitem[Noyes et al.(1984)]{noy84}
Noyes, R. W., Hartmann, L. W., Baliunas, S. L., Duncan, D. K., \& 
Vaughan, A. H., Rotation, Convection, and Magnetic Activity in Main-sequence
Stars, \textit{Astrophys. J.}, \textbf{279}, 763 - 777, 1984

\bibitem[Ray \& Heymann (1980)]{rh80}
Ray, J. \& Heymann, D., A Model for Nitrogen Isotopic Variations in the 
Lunar Regolith: Possible Solar System Contributions from a Nearby Planetary 
Nebula, \textit{Proc. Conf. Ancient Sun}, 491 - 512, 1980

\bibitem[Ribas et al.(2005)]{rib05}
Ribas, I., Guinan, E. F., G\"udel, M., Audard, M., 
Evolution of the Solar Activity over Time and Effects on Planetary 
Atmospheres. I. High-Energy Irradiances (1-1700\o{A})., 
\textit{Astrophys. J.}, \textbf{622}, 680 - 694, 2005

\bibitem[Saar (2001)]{saa01}
Saar, S. H., Recent Measurements of (and Inferences about) Magnetic 
Fields on K and M stars, \textit{the 11th Cool Stars, Stellar Systems 
and the Sun, ASP Conf. Series}, \textbf{223}, 292 - 299, 2001

\bibitem[Saar \& Brandenburg (1999)]{sb99}
Saar, S. H. \& Brandenburg, A., The Evolution of the Magnetic Activity 
Cycle Period. II. Results for an Expanded Stellar Sample, 
\textit{Astrophys. J.}, \textbf{524}, 295 - 310, 1999

\bibitem[Sakao et al.(2007)]{sak07}
Sakao, T. et al., Continuous Plasma Outflows from the Edge of a Solar 
Active Region as a Possible Source of Solar Wind., \textit{Science}, 
\textbf{318}, 1585 - 1588, 2007 

\bibitem[Stereborg et al.(2011)]{ste11}
Stereborg, M. G., Cohen, O., Drake, J. J., Gombosi, T. I., 
Modeling the Young Sun's Solar Wind and Its Interaction with Earth's 
Paleomagnetosphere, \textit{J. Geophys. Res.}, \textbf{116}, A01217, 2011  

\bibitem[Suzuki (2006)]{suz06}
Suzuki, T. K.,
Forcasting Solar Wind Speeds, 
\textit{Astrophys. J. Lett.}, \textbf{640}, L75 - L78, 2006

\bibitem[Suzuki (2007)]{suz07}
Suzuki, T. K., Structured Red Giant Winds with Magnetized Hot Bubbles 
and the Corona/Cool Wind Dividing Line, \textit{Astrophys. J.}, 
\textbf{659}, 1592 - 1610

\bibitem[Suzuki \& Inutsuka (2005)]{szi05}
Suzuki, T. K. \& Inutsuka, S.:
Making the corona and the fast solar wind: a self-consistent 
simulation for the low-frequency \Alfven waves from photosphere to 0.3AU, 
\textit{Astrophys. J. Lett.}, \textbf{632}, L49 - L52, 2005 

\bibitem[Suzuki \& Inutsuka (2006)]{si06}
Suzuki,T. K. \& Inutsuka, S.:Solar Winds Driven by Nonlinear Low-Frequency 
\Alfven Waves from the Photosphere : Parametric Study for Fast/Slow Winds 
and Disappearance of Solar Winds,
\textit{J. Geophys. Res.}, \textbf{111}, A6, A06101, 2006 (SI06)

\bibitem[Telleschi et al.(2005)]{tel05}
Telleschi, A. G\"udel, M., Briggs, K., Audard, M., Ness, J.-U., Skinner, S. L., 
Coronal Evolution of the Sun in Time: High-Resolution X-Ray Spectroscopy 
of Solar Analogs with Different Ages, \textit{Astrophys. J.}, \textbf{622}, 
653 - 679, 2005

\bibitem[Wang \& Sheeley (1990)]{ws90}
Wang, Y.-M. \& Sheeley, Jr, N. R., 
Solar wind speed and coronal flux-tube expansion, 
\textit{Astrophys. J.}, \textbf{355}, 726 - 732, 1990 

\bibitem[Wang \& Sheeley (1991)]{ws91}
Wang, Y.-M. \& Sheeley, Jr, N. R.,  
Why fast solar wind originates from slowly expanding coronal flux tubes, 
\textit{Astrophys. J. Lett.}, \textbf{372}, L45 - L48, 1991 

\bibitem[Weber \& Davis (1967)]{wd67}
Weber, E. J., Davis, L.,Jr. The Angular Momentum of the Solar Wind, 
\textit{Astorphys. J.}, \textbf{148}, 217 - 227, 1967

\bibitem[Wood et al.(2002)]{wod02}
Wood, B. E., M\"uller, H.-R., Zank, G. P., Linsky, J. L.,
Measured Mass-Loss Rates of Solar-like Stars as a Function of Age 
and Activity, \textit{Astorphys. J.}, \textbf{574}, 412 - 425, 2002

\bibitem[Wood et al.(2005)]{wod05}
Wood, B. E., M\"uller, H.-R., Zank, G. P., Linsky, J. L., Redfield, S., 
New Mass-Loss Measurements from Astrospheric Ly$\alpha$ Absorption, 
\textit{Astrophys. J. Lett.}, \textbf{628}, L143 - L146, 2005 

%% Format for Journal Reference
(Author), (article's title), \textit{(Journal)}, \textbf{(Volume)}, (page number)--(page number), (year).

%% Format for Book
%(Author), (article's title), in \textit{(Book's title)}, Edited by (Editor), (total page) pp,
%(publisher), (published place), (year).
%\end{references}
\end{thebibliography}
\end{document}